\documentclass[preprint,amsmath,amssymb,aip,superscriptaddress,nofootinbib]{revtex4-2}
\usepackage{amsmath}
\usepackage{natbib}
\usepackage{graphicx}
\usepackage{mathtools,accents}

\begin{document}

\newtheorem{lemma}{Lemma}
\newtheorem{corollary}{Corollary}

\title{The Effect of Pressure Fluctuations on the Shapes of Thinning Liquid Curtains}

\author
 {
 Bridget M. Torsey }\affiliation{School of Mathematical Sciences, Rochester Institute of Technology, Rochester, NY 14623, USA},
 \author
 {Steven J. Weinstein} \affiliation{School of Mathematical Sciences, Rochester Institute of Technology, Rochester, NY 14623, USA} \affiliation{Department of Chemical Engineering, Rochester Institute of Technology, Rochester, NY 14623, USA}, \author
 {David S. Ross} \affiliation{School of Mathematical Sciences, Rochester Institute of Technology, Rochester, NY 14623, USA},
 \author
 {Nathaniel S. Barlow} \affiliation{School of Mathematical Sciences, Rochester Institute of Technology, Rochester, NY 14623, USA}

\begin{abstract}
We consider the time-dependent response of a gravitationally-thinning inviscid liquid sheet (a coating curtain) leaving a vertical slot to sinusoidal ambient pressure disturbances. The theoretical investigation employs the hyperbolic partial differential equation developed by \cite{weinp1}. The response of the curtain is characterized by the slot Weber number, $W_{e_0} = \rho q V/2\sigma$, where $V$ is the speed of the curtain at the slot, $q$ is the volumetric flow rate per unit width, $\sigma$ is the surface tension, and $\rho$ is the fluid density. Flow disturbances travel along characteristics with speeds relative to the curtain of $\pm \sqrt{uV/W_{e_0}}$, where $u = \sqrt{V^2 + 2gx}$ is the curtain speed at a distance $x$ downstream from the slot. When the flow is subcritical ($W_{e_0} < 1$), upstream traveling disturbances near the slot affect the curtain centerline, and the slope of the curtain centerline at the slot oscillates with an amplitude that is a function of $W_{e_0}$. In contrast, all disturbances travel downstream in supercritical curtains ($W_{e_0} > 1$) and the slope of the curtain at the slot is vertical. Here, we specifically examine the curtain response under supercritical and subcritical flow conditions near $W_{e_0} = 1$ to deduce whether there is a substantial change in the overall shape and magnitude of the curtain responses. Despite the local differences in the curtain solution near the slot, we find that subcritical and supercritical curtains have similar responses for all imposed sinusoidal frequencies.
\end{abstract}

\maketitle

\section{Introduction}
\begin{figure}
\centering
  \includegraphics[width=8.5cm,height=12cm]{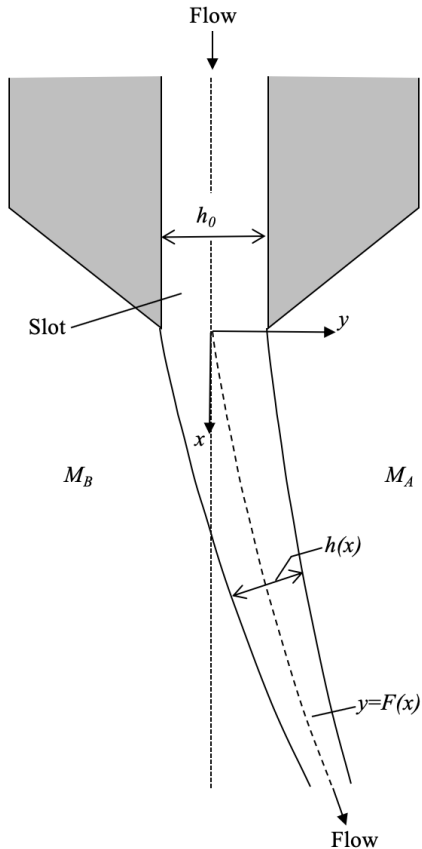}
  \caption{Side view schematic of a liquid exiting a slot of height $h_0$ and falling under the influence of gravity while subjected to ambient gas pressures $M_A$ and $M_B$ on its sides. The curtain is assumed to be infinite and invariant in the $z$ direction, oriented out of the figure. The centerline of the curtain and its local thickness are denoted as $y=F(x)$ and $h(x)$, respectively. The governing equations (1.1)-(1.3) are valid for curtain deflections $F<<h$, so dimensions in the figure are chosen for clarity and are not to drawn to scale.}
  \label{CurtainDiagram}
\end{figure}

 Curtain coating is a common industrial process that uses wide, thin planar liquid sheets (curtains) to deposit uniform thin films on moving substrates \citep{annual}. In one of its simplest configurations, a curtain leaves an inverted slot die and thins under the influence of gravity as it falls, see figure \ref{CurtainDiagram}. Curtains are subjected to ambient disturbances that deflect them; these deflections can cause nonuniform liquid coatings and imperfections in final dried products. Flow disturbances are often examined using linear theory because coated product quality is sensitive to even small thickness variations. Curtains have significant surface area in contact with the surrounding air along their long and wide faces, so understanding how they respond to pressure disturbances is of practical importance. In this study we focus on the effects of sinusoidal time-varying pressure disturbances on the shapes and deflections of curtains.  

 The response of liquid curtains to pressure disturbances has been well-studied both experimentally and theoretically. Much of that work has focused on the related problem of water bells, i.e. radially-symmetric liquid sheets (among the many references, see \cite{Hopwood}; \cite{Lance}; \cite{Ramos_1988}, \cite{Brunet_2004}, \cite{Paramati_2016}). In aggregate, this work shows that agreement with experiment is obtained when models use the approximation of inviscid flow with small variations in thickness about the curved centerline of the water bell; that is, the curtain is gradually thinning in the direction of flow. These studies also demonstrate that small pressure differences across the surface of a liquid sheet can have a large impact on the shape of water bells.   

 Curtains have been similarly studied with respect to ambient pressure disturbances. In previous work, \cite{weinp1} justify a potential flow approximation to the flow in a curtain, and derive an equation that governs deflections of a curtain's centerline, $y = F(x)$ shown in figure 1, given as:

\begin{equation}\label{eq1}
  \left(\frac{\partial}{\partial t}+g\frac{\partial}{\partial u}\right)^2 F - \frac{2\sigma g^2}{\rho q}\frac{\partial}{\partial u}\left(\frac{1}{u}\frac{\partial F}{\partial u}\right) =\frac{(M_B-M_A)u}{\rho q}
\end{equation}
where:
\begin{equation} \label{eq2a}
  u = \sqrt{V^2+2gx}
\end{equation}
and the local thickness of the curtain, $h$, is expressed as:
\begin{equation} \label{eq2b}
  h(x) = \frac{q}{u}.
\end{equation}

 In (\ref{eq1}) - (\ref{eq2b}), $t$ is time, $u$ is the curtain speed at position $x$, $V$ is the curtain speed at the slot, $g$ is the acceleration of gravity, $q$ is the volumetric flow rate, $\sigma$ is the surface tension (which we take to be constant), $\rho$ is the liquid density, and $M_A$ and $M_B$ are the respective pressures on the front and back faces of the curtain as shown in figure \ref{CurtainDiagram}. \cite{weinp1} derive the equation (1.1) as follows. The equation for the velocity field and thickness of the undisturbed time-independent curtain (centerline $y=0$ for all $x$) is determined via an asymptotic expansion in the small parameter $\epsilon = gq/V^3$. The limit $\epsilon \rightarrow 0$ corresponds physically to a gradually-thinning liquid curtain. The resulting lowest-order flow in the curtain is plug at each location $x$ according to equation (1.2), where velocity variations across the curtain thickness (i.e. the $y$-direction in figure \ref{CurtainDiagram}) are of $\mathcal{O}(\epsilon^2)$. The time-dependent potential flow equations are then linearized for small perturbations about the asymptotic steady-state equations (the base flow). Since the base flow is approximate, terms to $\mathcal{O}(\epsilon^2)$ are required to preserve accuracy in the linearization. The resulting time-dependent equation (1.1) is valid for curtain deflections such that $F<<h$. In the limit taken, pressure disturbances do not affect the local thickness of the curtain to leading order, so the resulting curtain response is captured by the deflection of its centerline \citep{weinp1}.    

 When examined under conditions of small deflection, the equation of \cite{finn} is identical to (1.1) at steady state. Equations (\ref{eq2a}) and (\ref{eq2b}) are valid regardless of the magnitude of deflection provided that the curtain is long and thin. \cite{weinp1} shows via dominant balance that the inviscid result, equation (1.2), is consistent with Taylor’s equation (derived in the Appendix of \cite{Brown_1961}) for large $x$, even when viscosity is included. Predictions of (\ref{eq1}) and (1.2) agree well with experiments under steady-state (\cite{finn}) and transient (\cite{clarkep2}) conditions. In the latter study, \cite{clarkep2} demonstrate that predictions agree with experiments when the initial velocity is adjusted to account for the effect of an entrance region through extrapolation to the slot location at $x=0$. This result agrees with the empirical equation of \cite{Brown_1961} for an undisturbed vertical curtain. Note that equations (1.1) and (1.2) are strictly valid in a region displaced downstream from the slot because the loss in viscous traction from the slot leads to a flow rearrangement not captured in these equations (\cite{clarke1968}, \cite{Tillet_1968}; \cite{Ruschak_1980}; \cite{Georgiou_1988}).  Other studies have used the slender curtain/inviscid approach to examine the effect of nonlinear dynamics (\cite{ramos}) and stationary wave formation (\cite{deluca_1997}).    

 Closely related to the current work is the nappe oscillation configuration, which is used to model observed perturbations in liquid sheets formed as water flows over waterfalls, dams, and weirs.  In such a configuration, pressure disturbances are affected by the curtain motion via an enclosure that includes the curtain as one of its long and wide sides; the other side of the curtain is maintained at atmospheric pressure.  As a result, the motion of the curtain affects the volume of the enclosed region, which affects the pressure in that region and provides a restoring force.  There have been many experimental studies of this phenomena (see for example, \cite{Binnie_1974}; \cite{Sato_2007}; \cite{Mori_2012}), and observed oscillation frequencies measured in the curtain correlate with those measured in the enclosure itself.  Theoretical analyses follow the modeling approach used to obtain equations (1.1) to (1.3) and predict the natural frequencies of the system in the absence of surface tension (\cite{Schmid_2002}, \cite{derosa_2014}); these provide a more precise model of the curtain dynamics than the simplified theory provided by \cite{Binnie_1974}.  The more recent analysis of \cite{girf} incorporates surface tension and predicts natural frequencies that agree favorably with the cited experiments of \cite{Binnie_1974} and \cite{Sato_2007}.   

 The governing equation, (1.1), is a second-order hyperbolic partial differential equation (PDE). The two sets of characteristics associated with the PDE carry information that moves relative to the curtain at speeds of $\pm \sqrt{uV/W_{e_0}}$. It is established in hyperbolic PDE theory that the number of constraints specified along a boundary must be equal to the number of characteristics that emanate from that boundary at each point \citep{lax}. The directions of the characteristics, and therefore the associated boundary conditions, are determined by the Weber number at the slot, given by $ W_{e_0}= \rho q V/2\sigma$. In supercritical curtains ($W_{e_0} > 1$), both sets of characteristics leave the slot and are oriented downstream. Thus, two conditions must be specified along this boundary. This constraint placement corresponds to the physics of supercritical flows, in which the momentum flux is greater than the surface tension at every location, so any disturbances are washed downstream. \cite{finn} establish experimentally that supercritical curtains leave the slot vertically under steady conditions, confirming that points downstream do not influence upstream locations. In a subcritical curtain ($ W_{e_0} < 1$), there is a region in which one set of characteristics is oriented upstream and the other set is oriented downstream. This region begins at the slot and ends at the critical point, the point in the curtain at which the local Weber number, $ W_e = \rho q u/2\sigma$, equals 1. Downstream of the critical point, both sets of characteristics are oriented downstream. Therefore, if $ W_{e_0} < 1$, only one set of characteristics leaves the slot, so only one condition is applied along this boundary. Disturbances in this region of the curtain move upstream and downstream along the characteristics. \cite{finn} demonstrate experimentally that when a constant pressure drop is applied across a curtain containing such a region, the curtain takes on an angle at the slot. They show theoretically that this angle can be predicted by removing a singularity in the governing equation that arises at the location where $W_e = 1$, the precise location where the direction of characteristics changes. As demonstrated in the appendix of \cite{finn}, a subcritical water bell issuing from an annular slit and having an applied pressure difference takes on an angle at the slit. Their analytical solution, obtained under conditions of negligible gravity, accounts for the earlier qualitative experimental observations of \cite{Baird_1962}. \cite{Ramos_1997} confirms this conclusion in a later examination of this problem. \cite{Brunet_2004} examine water bells with applied pressure differences that undergo a subcritical-to-supercritical transition, both experimentally and theoretically. They predict water bell shapes that agree well with experiment. As noted previously, the water bell equations use the same thin-film-modeling assumptions that \cite{finn} used to study the planar configuration, so these water bell results also may be used to confirm the gradually-thinning and inviscid assumptions used in the development of equations (1.1) to (1.3).   

 With this background, we now return to the nappe configuration analysis of \cite{girf} cited above. There, the equation governing the curtain motion caused by the applied pressure is the same as that of equation (1.1) although the pressure disturbances $M_A-M_B$ are expressed in terms motion of the curtain shape itself through the enclosure volume. Despite this mathematical difference, the characterization of subcritical and supercritical flows discussed above in terms of $W_{e_0}$ applies. \cite{girf} examine curtain flows with slot Weber numbers near one. As we discussed above, when the flow is subcritical ($W_{e_0}<1$), there exists a location in the curtain at which $W_e=1$, below which the flow becomes supercritical. As in the steady problem of \cite{finn}, the time-dependent equation is singular at the point at which $W_e=1$. Because (1.1) is of second order in the spatial dimension, the general solution of the initial-boundary value problem has two degrees of freedom that allow us to satisfy spatial constraints. In supercritical cases we use those degrees of freedom to specify two conditions: that the curtain centerline is not displaced at the slot exit, and that the curtain leaves the slot with its centerline aligned with that of the slot. In subcritical cases we still use one degree of freedom to enforce the condition that the curtain centerline is not displaced at the slot exit, but we use the second degree of freedom to enforce the condition that the solution be smooth at the singular point. This switch in the application of the second degree of freedom corresponds to the physical facts that subcritical curtains leave the slot at non-zero, and sometimes varying, angles, and that the transition from subcritical to supercritical flow in such curtains is smooth. In this paper we refer to subcritical curtains as those for which $W_{e_0}<1$ and for which there exists a location where the flow transitions from subcritical to supercritical. This definition is necessary, as \cite{finn} show experimentally that it is possible for curtains to remain subcritical over their entire lengths. This observation is also consistent with experimental findings of \cite{deluca_1999}, who show that long subcritical curtains can persist without rupture. We define supercritical curtains as those for which $W_{e_0}>1$ and thus the flow is supercritical over the entire domain. For supercritical curtains, \cite{girf} and \cite{finn} both find that the slope of the curtain at the slot is vertical.

 \cite{girf} report a distinct increase in dominant oscillation frequency as the slot Weber number is decreased from supercritical to subcritical. It is not clear from the investigation whether this effect is a result of an inherent susceptibility of the curtain to pressure disturbances because of the oscillation of the curtain centerline slope at the slot exit, or is a result of the pressure coupling present in a nappe configuration. In this paper we study the imposition of a pressure drop across the curtain that is sinusoidal in time and is not coupled to the curtain motion. This configuration is itself practically relevant to curtain coating processes. We specifically examine the response of the curtain as the flow is reduced from supercritical to subcritical, and we determine whether the corresponding change in the boundary conditions at the slot exit leads to a substantial change in the overall shape and magnitude of the curtain response. This provides a comparison of the capacities of subcritical and supercritical curtains to resist pressure disturbances.

\section{Theory}

 The dimensionless form of (\ref{eq1}) is

\begin{equation} \label{dimensionless}
\frac{\partial^2\bar{F}}{\partial\bar{t}^2}+ 2\bar{u} \frac{\partial^2\bar{F}}{\partial\bar{x}\partial\bar{t}}+ \bar{u}\frac{\partial}{\partial\bar{x}}\left[\left(\bar{u}-\frac{1}{W_{e_0}}\right)\frac{\partial\bar{F}}{\partial\bar{x}}\right]=\bar{u}(\bar{P}_2-\bar{P}_1).\end{equation} 

 Here, $\bar{P}_2-\bar{P}_1$ is the pressure difference across the curtain, whose displacement from its centerline is $\bar{F}(\bar{x},\bar{t})$ \citep{weinp1}. The dimensionless version of (1.2) is 

\begin{equation}
\bar{u} = \sqrt{1+2\bar{x}}, \end{equation}

 where dimensionless variables are defined as:

\refstepcounter{equation}
$$
  \bar{F} = \frac{F}{h_0} \qquad \bar{u} = \frac{u}{V} \qquad \bar{x} = \frac{xg}{V^2} \qquad \bar{t} = \frac{tg}{V}
\eqno{(\theequation{\mathit{a},\mathit{b},\mathit{c},\mathit{d}})}
$$
\refstepcounter{equation}
$$\bar{P}_2-\bar{P}_1 = \frac{M_B-M_A}{\rho \epsilon^2 V^2} \qquad \epsilon = \frac{gq}{V^3} \qquad W_{e_0} = \frac{\rho q V}{2\sigma}.
\eqno{(\theequation{\mathit{a},\mathit{b},\mathit{c}})}
$$
 In (2.3a), $h_0$ is the slot height, and other variables have been defined in section 1 and shown schematically in figure 1. We then convert the coordinate system from Eulerian to Lagrangian by making the following variable change:

\begin{equation}
  \xi = \bar{u} - 1.
\end{equation}

 In order to study the response of a curtain, we consider an applied sinusoidal pressure disturbance of the form $\bar{P}_2-\bar{P}_1= \beta e^{i \bar{\omega} \bar{t}}$ such that (2.1) is written, using coordinate transformation (2.5), as:

\begin{equation} \label{PDE}
  \frac{\partial^2\bar{F}}{\partial\bar{t}^2}+ 2 \frac{\partial^2\bar{F}}{\partial\xi\partial\bar{t}}+ \frac{\partial}{\partial\xi}\left(\left(1-\frac{1}{(\xi +1)W_{e_0}}\right)\frac{\partial\bar{F}}{\partial\xi}\right)=(\xi +1)\beta e^{i \bar{\omega} \bar{t}}
\end{equation}

 where $\bar{\omega}$ and $\beta$ are given by

\refstepcounter{equation}
$$
\bar{\omega}= \frac{\omega V}{g} \qquad \beta = \frac{\alpha e^{i\theta_R}}{\rho \epsilon^2 V^2}.
\eqno{(\theequation{\mathit{a},\mathit{b}})}
$$

 Here, $\omega$ is the angular frequency and $\alpha$ and $\theta_R$ are the magnitude and reference phase of the pressure disturbance, respectively. The reference phase, $\theta_R$, is adjusted to provide clarity in the presentation of results to follow (see discussion in Section 3). It is understood that only the real part of $\bar{F}$ is taken as the actual solution. The periodic solution for $\bar{F}$ is of the form

\begin{equation} \label{F}
  \bar{F}(\xi,\bar{t}) = \beta \bar{H}(\xi)e^{i\bar{\omega}\bar{t}}
  \end{equation}
  
 where $\bar{H}(\xi)$ is to be determined. We obtain a second-order ordinary differential equation (ODE) by substituting (\ref{F}) into (\ref{PDE}):
\begin{equation} \label{ODE1}
\left((\xi +1)^2-\frac{(\xi +1)}{W_{e_0} }\right)\frac{d^2\bar{H}}{d \xi^2}+\left(2i(\xi +1)^2\bar{\omega}+\frac{1}{W_{e_0} }\right)\frac{d\bar{H}}{d \xi}-(\xi +1)^2\bar{\omega}^2\bar{H}=(\xi +1)^3.
\end{equation}

 A second change of coordinates is implemented to simplify calculations, given by:

\refstepcounter{equation}
$$
z= \xi - c \qquad c = \frac{1}{W_{e_0} } - 1.
\eqno{(\theequation{\mathit{a},\mathit{b}})}
$$

 In these coordinates, the critical point is $z=0$ and (\ref{ODE1}) becomes

\begin{equation}\label{ODE}
z(z+c+1)\frac{d^2\bar{H}}{dz^2}+(2i(z+c+1)^2\bar{\omega}+c+1)\frac{d\bar{H}}{dz}-(z+c+1)^2\bar{\omega}^2\bar{H}=(z+c+1)^3.
\end{equation}

 As discussed in section 1, there are two characteristics that are oriented downstream from the slot in supercritical curtains ($ W_{e_0} > 1$), and the appropriate constraints to apply here are

\begin{equation} \label{IC1}
  \bar{H}(-c) = 0
\end{equation}

\begin{equation} \label{IC2}
  \frac{d\bar{H}}{dz}(-c) = 0.
\end{equation}   

 Note that the subcritical case ($W_{e_0} < 1$) corresponds to $c>0$. Condition (\ref{IC1}) sets the curtain centerline to be that of the slot, and (\ref{IC2}) specifies that the curtain be vertical there. We can see that for supercritical curtains, (\ref{ODE}), along with the conditions stated in (\ref{IC1}) and (\ref{IC2}), constitute a well-posed initial value problem \citep{john}. Supercritical solutions for this system may be obtained over the entire domain using a fourth-order accurate Runge-Kutta marching scheme.   

 For subcritical curtains ($ W_{e_0} < 1$), only one characteristic is oriented downstream from the slot, so we only apply (\ref{IC1}). Furthermore, the coefficient of the highest order term in  (\ref{ODE}) goes to zero at the critical point ($z=0$), and therefore the ODE is singular (\cite{hildebrand}). An implication of this singularity is that any marching scheme will become infinitely stiff as $z$ approaches 0. Thus, we use a power series centered at $z=0$ to determine the solution from the slot ($z = -c$) to a point downstream from the transition point ($z = c$). Then, taking the first coefficients of the power series solution as initial conditions, we use the Runge-Kutta scheme mentioned above to complete the solution from $z = c$ to the end of the curtain ($z = z_L$); the flow is supercritical below the critical point, allowing us to apply a marching scheme in this way. We present the derivation of the power series in appendix \ref{AppendixPower}. In the power series, enough terms are used to achieve machine precision.   

\section{Results and Discussion}

 Figures 2-9 provide results for subcritical and supercritical curtains corresponding to the configuration shown in figure 1.  Note that these figures have been rotated counter-clockwise 90 degrees, such that the gravitational force points right, along the $\bar{x}$-axis. In all of the figures, the location of the unperturbed curtain centerline, $\bar{y}=0$, is shown as a dashed line and the shapes that the curtains adopt at 1/4 increments of the periods of motion are represented by the solid lines. For clarity in the presentation of results, the reference phase of the pressure disturbance, $\theta_R$ in (2.7b), is chosen to be 
\begin{equation}
    \theta_R=\arctan \left(\frac{\text{Im}(\bar{H}(\bar{x}_L))}{\text{Re}(\bar{H}(\bar{x}_L))}\right),
  \end{equation}
  
 where $\text{Im}$ and $\text{Re}$ denote the imaginary and real parts of the complex argument. Note that reference phase merely adjusts the time at which the curtain adopts a shape, but does not affect the sequence of shapes predicted through its period of oscillation. Equation (3.1) assures that the bottom of the curtain, located at $\bar{x}=\bar{x_L}$, sweeps through the same locations as it moves forward and backward throughout an oscillation cycle--this makes it easier to interpret the curtain response.  

 Figures 2 and 3 provide curtain responses for $W_{e_0}=1.1$ (supercritical) and $W_{e_0} = 0.9$ (subcritical) conditions both with $\bar{\omega}=0.1$. The insets in these figures show the magnifications of the curtain shapes near the slot exit. As described earlier, the centerline slope of supercritical curtains is always zero at the slot exit, while the slope of supercritical curtains changes throughout the oscillation cycle. Despite this change in behavior, figures 2 and 3 show that the overall shape of the curtain is quite similar for subcritical and supercritical curtains, with the magnitude of the subcritical responses comparable. Figures 4 and 5 ($\bar{\omega}=0.5$)  and figures 6 and 7 ($\bar{\omega}=2$) compare $W_{e_0}=1.1$ and $W_{e_0}=0.9$ curtain responses. Again, the magnitude and shapes of the curtain responses are similar for supercritical and subcritical flows. Figures 8 and 9 provide additional results for $W_{e_0}=1.3$ and $W_{e_0}=0.7$, respectively, both with $\bar{\omega}=2$. Even though the range of Weber numbers has extended further into the supercritical and subcritical regimes compared with figures 6 and 7, there is no marked change in curtain shape or magnitude. Additionally, we have solved Equation 2.11 for a selection of subcritical and supercritical slot Weber numbers ($W_{e_0}$) sweeping over a wide range of frequencies. The trends shown in Figures 2-7 are maintained over all frequencies and slot Weber numbers surveyed. That is, the number of spatial oscillations within a given curtain length increases, and the maximum magnitude of curtain responses decreases with increasing frequency.   

 The latter observation is relevant as it mathematically eliminates the possibility of forcing the curtain via a pressure disturbance in such a way that natural curtain frequencies are excited. If resonance were to arise, there would be a large amplification in the vicinity of a given imposed frequency. At the precise resonant frequency, the assumed form of the forced solution (2.8) would become invalid, necessitating a secular time dependence (such as $t e^{\bar{\omega} t}$). The monotonic decrease in curtain deflection with increasing frequency (as seen in figures 2 through 9) demonstrates that such behavior does not occur. This mathematical result has a physical basis. In supercritical curtains, disturbances are washed downstream, so the reinforcement of repeated disturbances that characterizes resonance is not possible. In curtains that are subcritical near the slot but turn supercritical downstream, one might expect that some reinforcement is possible, because disturbances in the subcritical region propagate upstream to the slot. However these reflect at the slot and are washed downstream.   

 Figure \ref{end} provides the maximum curtain angle at the slot as a function of $W_{e_0}$ for various frequencies. Note that the curtain takes on a range of angles at the slot as it cycles through its motion. The maximum angle of the response at the slot immediately drops to zero as the slot Weber number increases past $W_{e_0} =1$, in accordance with a subcritical to supercritical transition.   

 Despite the marked difference in curtain motion at the slot quantified by figure 10, the subcritical and supercritical curtain responses shown in figures 2-9 are not appreciably different at a given frequency. These results indicate that there is not a significant change in a curtain's sensitivity to pressure disturbances in the subcritical to supercritical transition. The abrupt frequency shift observed by \cite{girf} in this transition is thus attributed to the coupling between the pressure disturbances and curtain motion that arises in the nappe configuration.

\section{Conclusion}
 We consider the time-dependent response of a gravitationally-thinning liquid curtain that is subjected to sinusoidal ambient pressure disturbances under subcritical and supercritical conditions. Consistent with previous studies under both steady state and transient conditions, we find that the centerline slope of the curtain is not the same as that of the slot under subcritical conditions.  This is a direct consequence of the flow of information propagating along characteristics in the governing hyperbolic PDE. We examine whether the increased susceptibility of the curtain to pressure disturbances arises because of the oscillation in curtain centerline slope at the slot exit for subcritical curtains. Our investigation shows there are no abrupt differences in the responses of subcritical and supercritical curtains for Weber numbers near 1 for all imposed sinusoidal frequencies. Experimental confirmation of these results is needed in future studies.

\clearpage
\begin{figure}
  \centering
  \includegraphics[width=10cm,height=8.25cm]{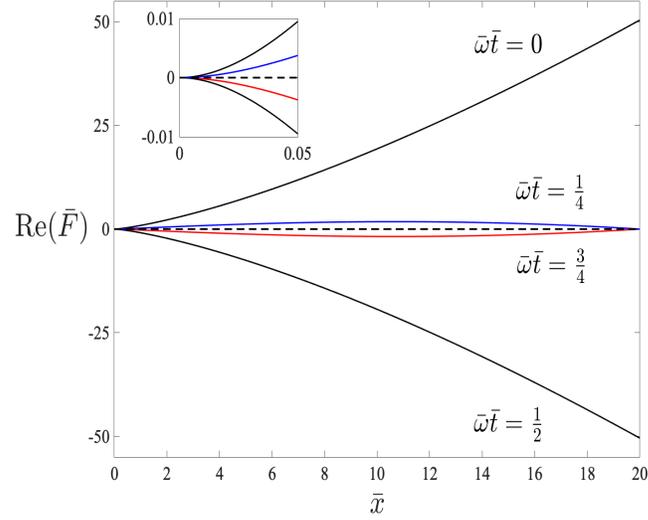}
  \caption{The location of the curtain centerline, $\text{Re}(\bar{F})$, as a function of distance down the curtain, $\bar{x}$, when it is subjected to a sinusoidal pressure disturbance. The inset is a magnification of the region near the slot. Here, the flow is supercritical ($W_{e_0}=1.1$) and $\bar{\omega}=0.1$, $\beta=1$, and $\theta_R=-0.4048$ radians (see (3.1)). The dashed line denotes the centerline of the unperturbed curtain.}
  \label{super1}
\end{figure}

\begin{figure}
  \centering
  \includegraphics[width=10cm,height=8.25cm]{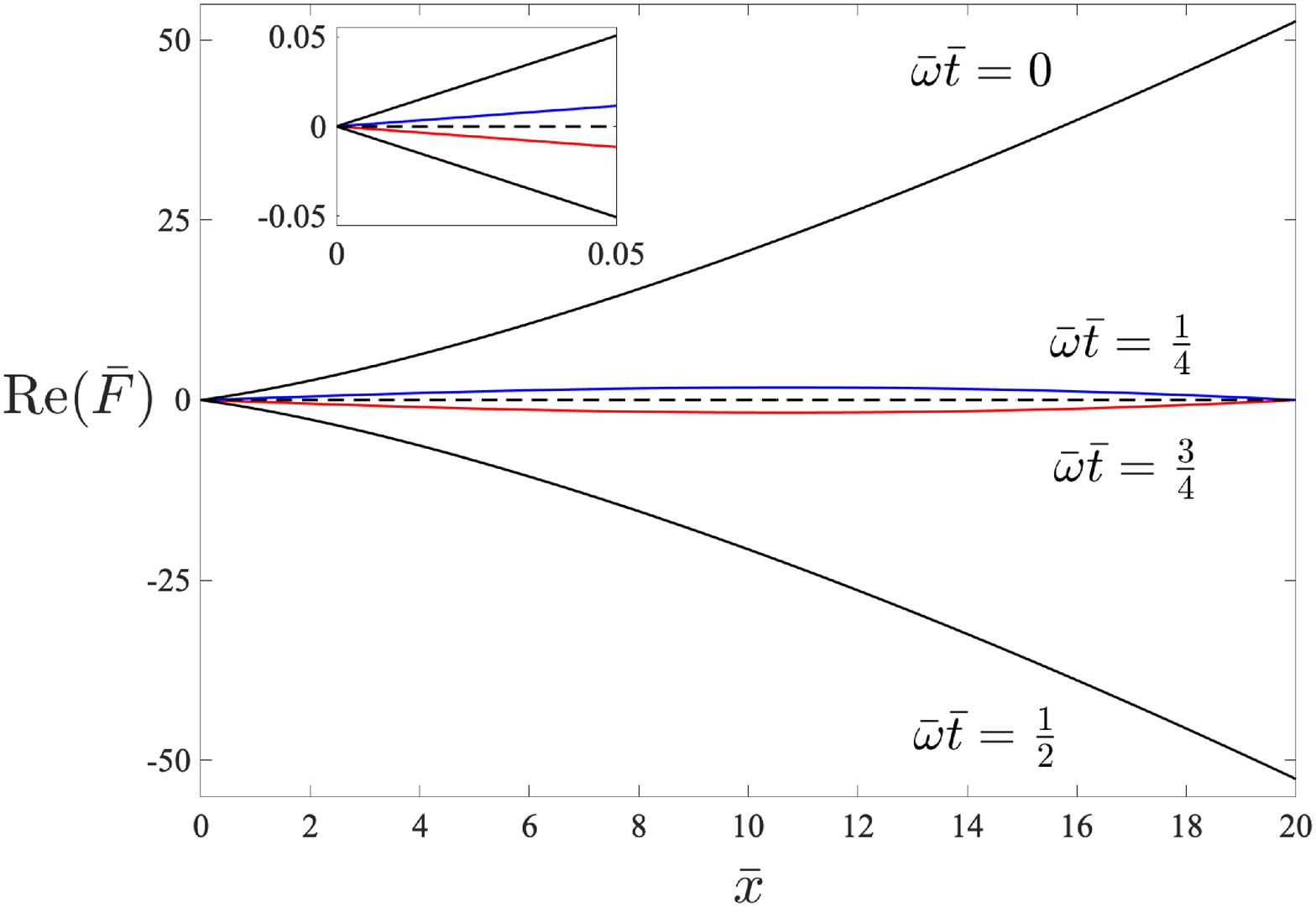}
  \caption{The location of the curtain centerline, $\text{Re}(\bar{F})$, as a function of distance down the curtain, $\bar{x}$, when it is subjected to a sinusoidal pressure disturbance. The inset is a magnification of the region near the slot. Here, the flow is subcritical ($W_{e_0}=0.9$) and $\bar{\omega}=0.1$, $\beta=1$, and $\theta_R=-0.4366$ radians (see (3.1)). The dashed line denotes the centerline of the unperturbed curtain.}
  \label{sub1}
\end{figure}

\begin{figure}
\centering
  \includegraphics[width=10cm,height=8.25cm]{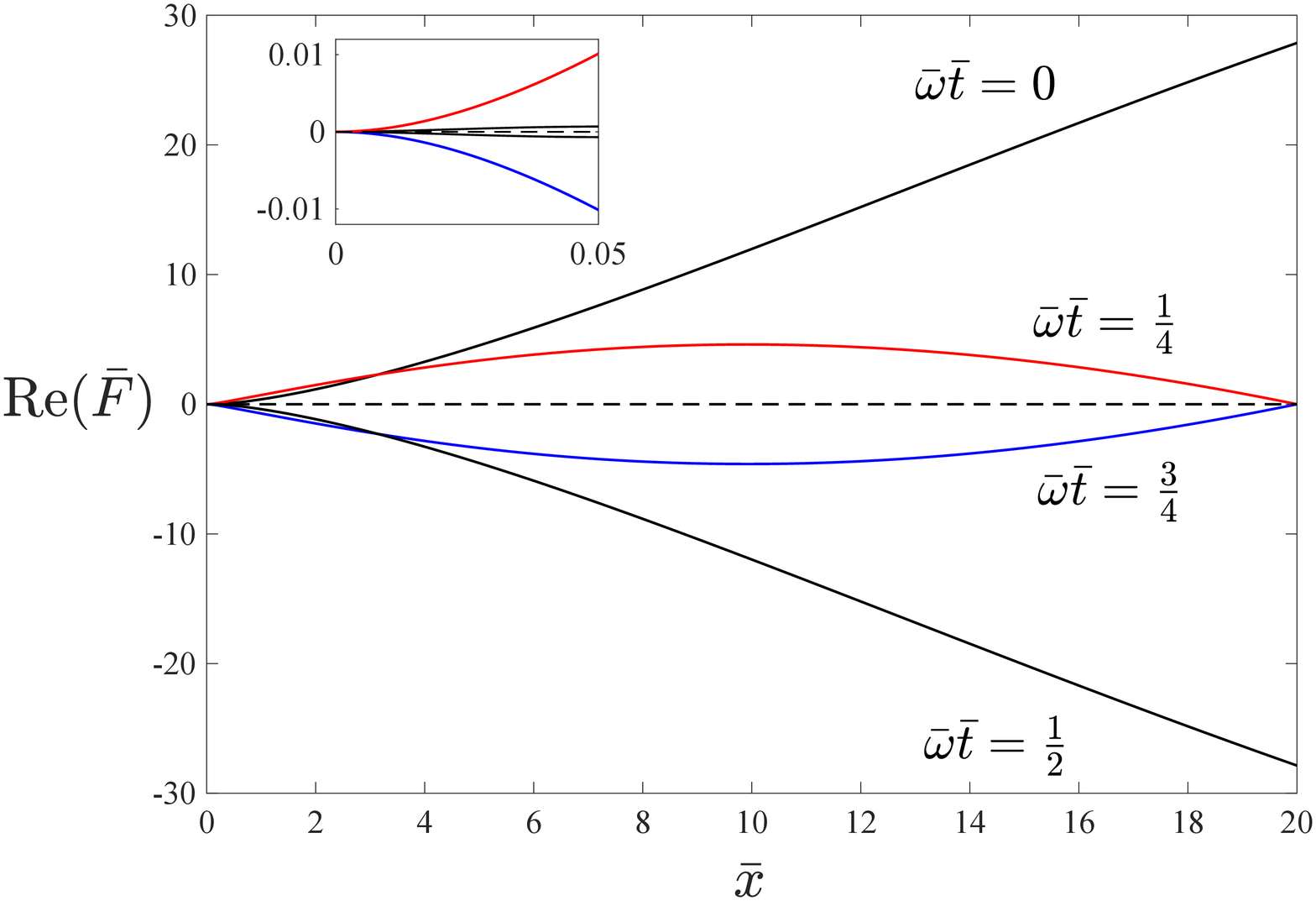}
  \caption{The location of the curtain centerline, $\text{Re}(\bar{F})$, as a function of distance down the curtain, $\bar{x}$, when it is subjected to a sinusoidal pressure disturbance. The inset is a magnification of the region near the slot. Here, the flow is supercritical ($W_{e_0}=1.1$) and $\bar{\omega}=0.5$, $\beta=1$, and $\theta_R=-1.7794$ radians (see (3.1)). The dashed line denotes the centerline of the unperturbed curtain.}
  \label{super2}
\end{figure}

\begin{figure}
  \centering
  \includegraphics[width=10cm,height=8.25cm]{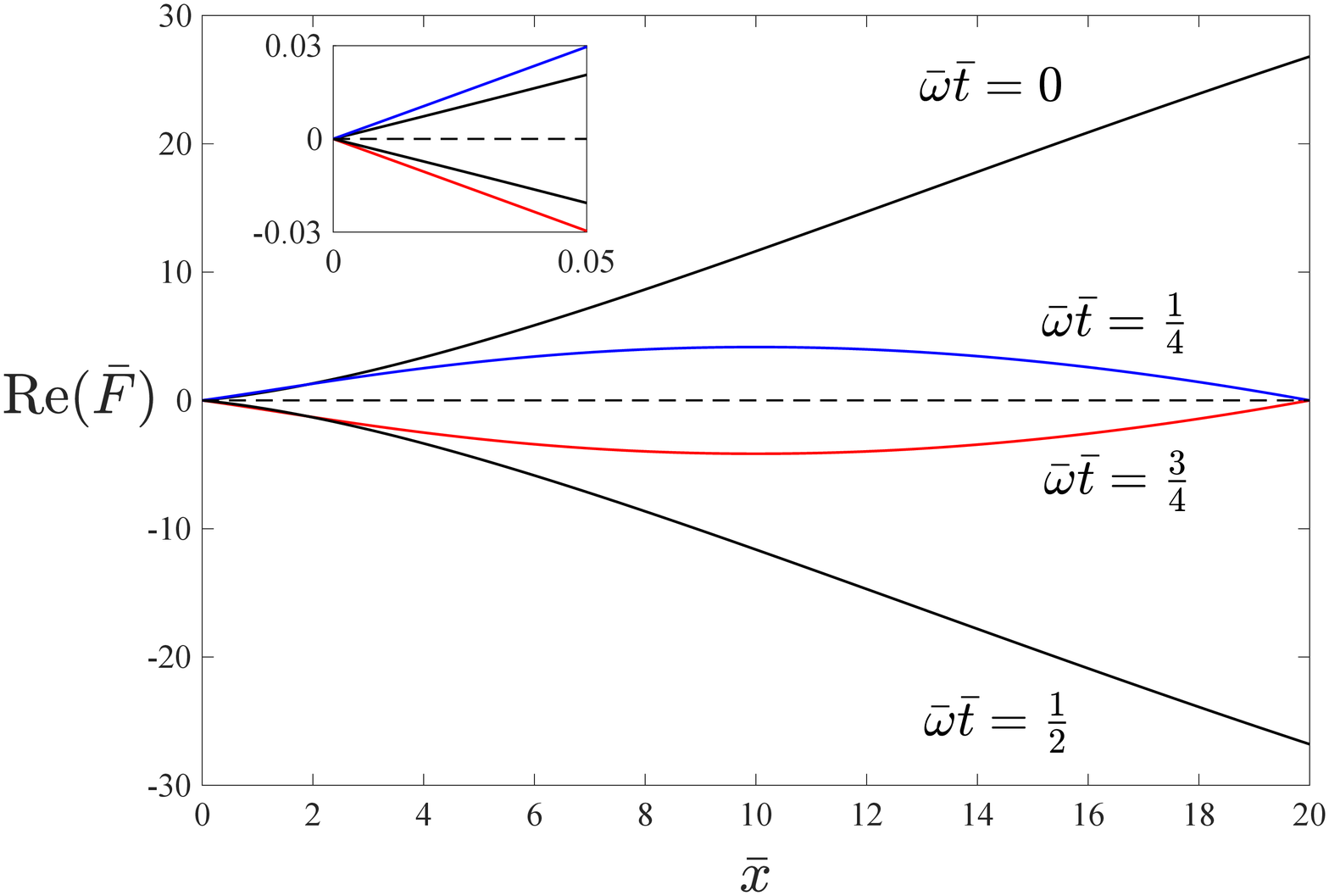}
  \caption{The location of the curtain centerline, $\text{Re}(\bar{F})$, as a function of distance down the curtain, $\bar{x}$, when it is subjected to a sinusoidal pressure disturbance. The inset is a magnification of the region near the slot. Here, the flow is subcritical ($W_{e_0}=0.9$) and $\bar{\omega}=0.5$, $\beta=1$, and $\theta_R=-1.7924$ radians (see (3.1)). The dashed line denotes the centerline of the unperturbed curtain.}
  \label{sub2}
\end{figure}

\begin{figure}
  \centering
  \includegraphics[width=10cm,height=8.25cm]{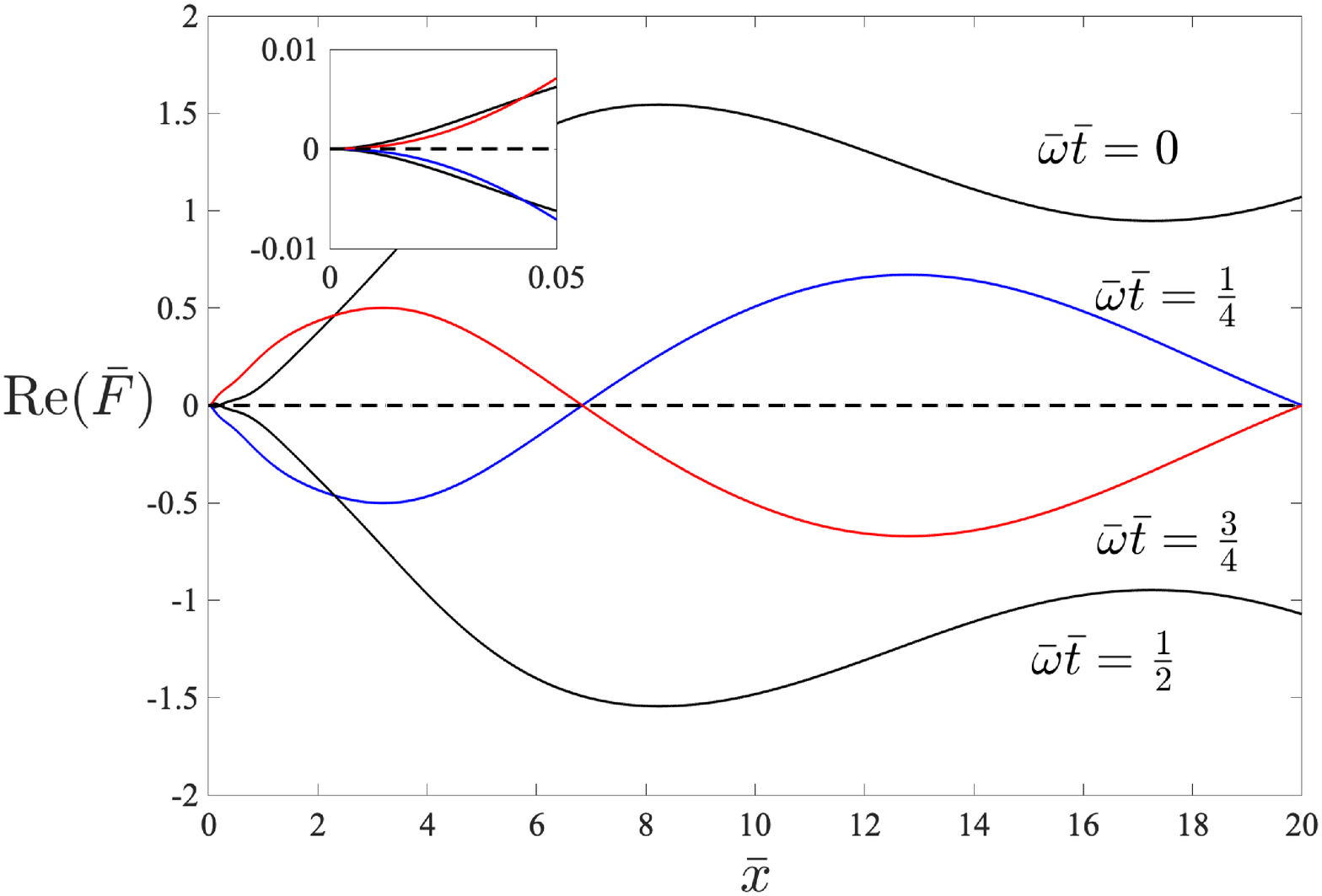}
  \caption{The location of the curtain centerline, $\text{Re}(\bar{F})$, as a function of distance down the curtain, $\bar{x}$, when it is subjected to a sinusoidal pressure disturbance. The inset is a magnification of the region near the slot. Here, the flow is supercritical ($W_{e_0}=1.1$) and $\bar{\omega}=2$, $\beta=1$, and $\theta_R=-2.8491$ radians (see (3.1)). The dashed line denotes the centerline of the unperturbed curtain.}
  \label{super4}
\end{figure}

\begin{figure}
  \centering
  \includegraphics[width=10cm,height=8.25cm]{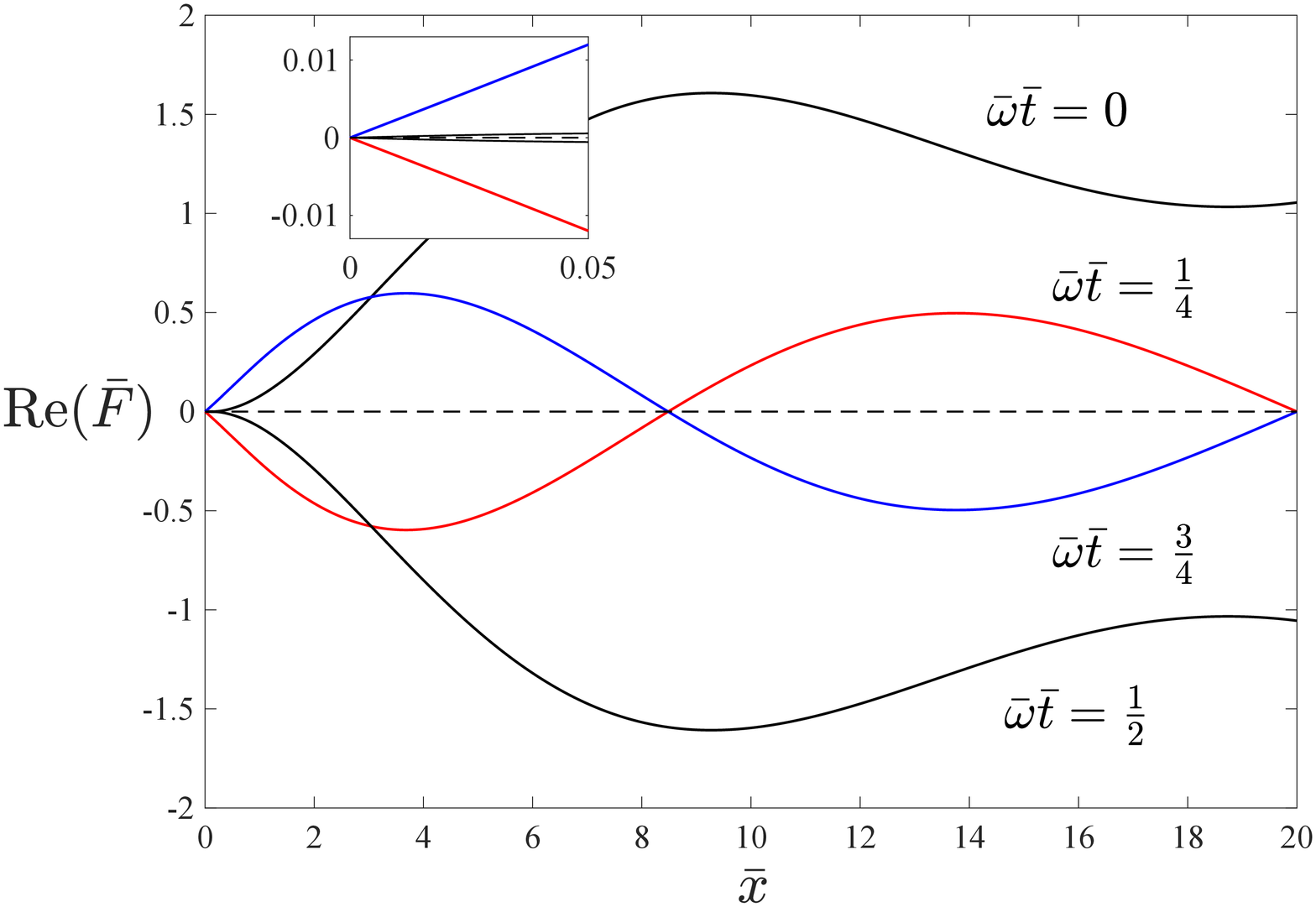}
  \caption{The location of the curtain centerline, $\text{Re}(\bar{F})$, as a function of distance down the curtain, $\bar{x}$, when it is subjected to a sinusoidal pressure disturbance. The inset is a magnification of the region near the slot. Here, the flow is subcritical ($W_{e_0}=0.9$) and $\bar{\omega}=2$, $\beta=1$, and $\theta_R=-2.9888$ radians (see (3.1)). The dashed line denotes the centerline of the unperturbed curtain.}
  \label{sub3}
\end{figure} 

\begin{figure}
\centering
  \includegraphics[width=10cm,height=8.25cm]{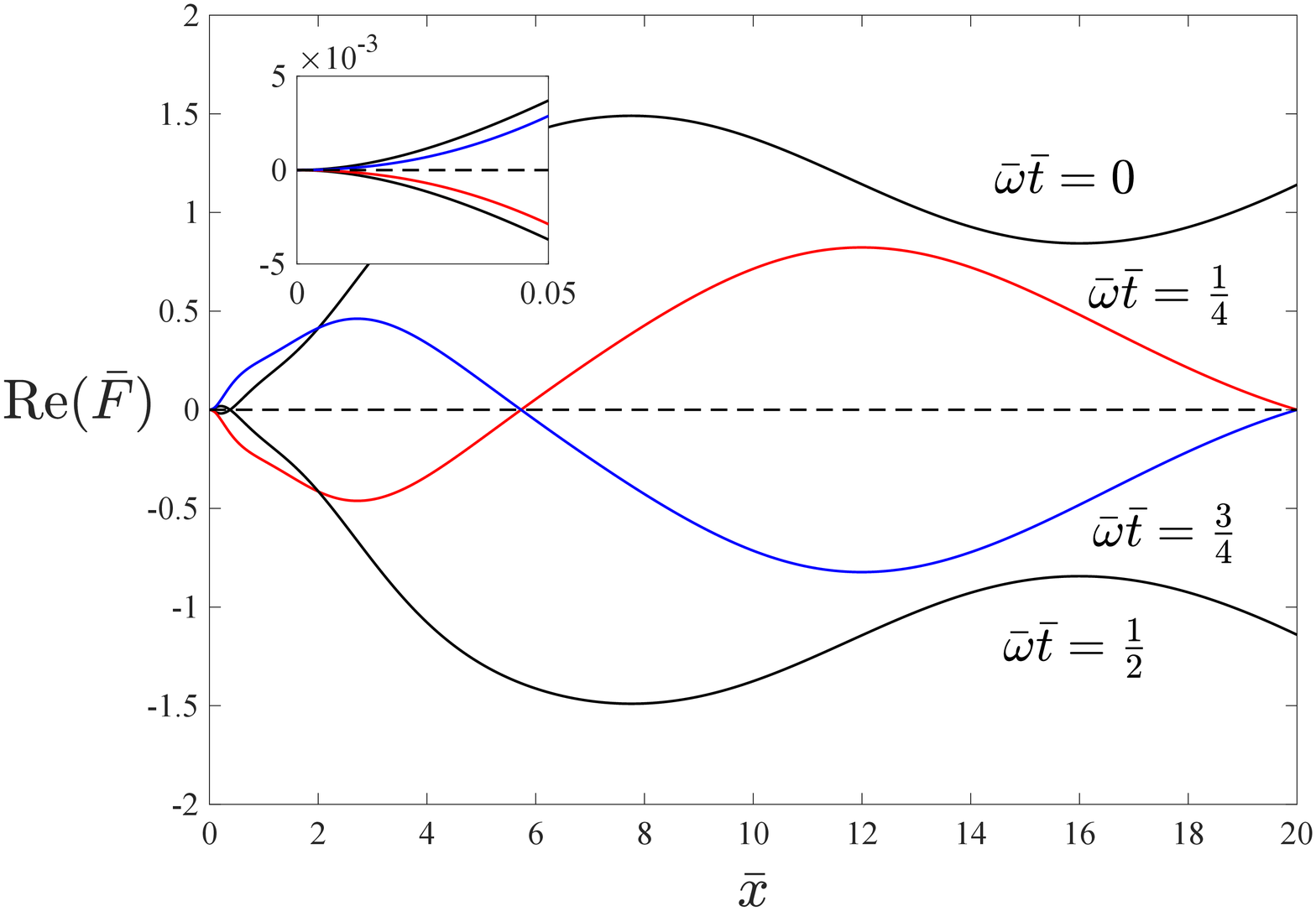}
  \caption{The location of the curtain centerline, $\text{Re}(\bar{F})$, as a function of distance down the curtain, $\bar{x}$, when it is subjected to a sinusoidal pressure disturbance. The inset is a magnification of the region near the slot. Here, the flow is supercritical ($W_{e_0}=1.3$) and $\bar{\omega}=2$, $\beta=1$, and $\theta_R=-2.7377$ radians (see (3.1)). The dashed line denotes the centerline of the unperturbed curtain.}
\end{figure}

\begin{figure}
  \centering
  \includegraphics[width=10cm,height=8.25cm]{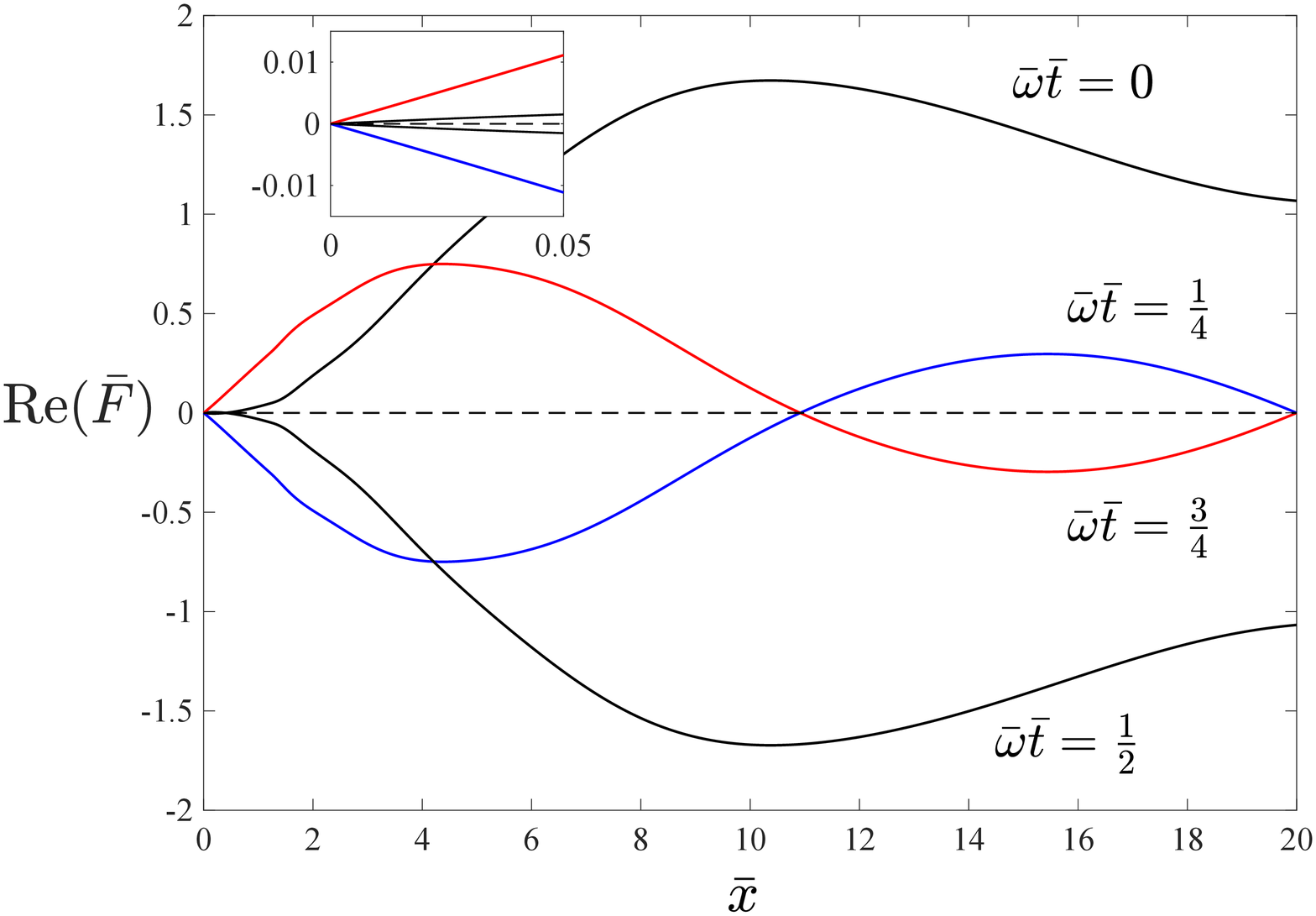}
  \caption{The location of the curtain centerline, $\text{Re}(\bar{F})$, as a function of distance down the curtain, $\bar{x}$, when it is subjected to a sinusoidal pressure disturbance. The inset is a magnification of the region near the slot. Here, the flow is subcritical ($W_{e_0}=0.7$) and $\bar{\omega}=2$, $\beta=1$, and $\theta_R=3.1252$ radians (see (3.1)). The dashed line denotes the centerline of the unperturbed curtain.}
\end{figure} 

\begin{figure}
  \centering
  \includegraphics[width=11cm,height=8cm]{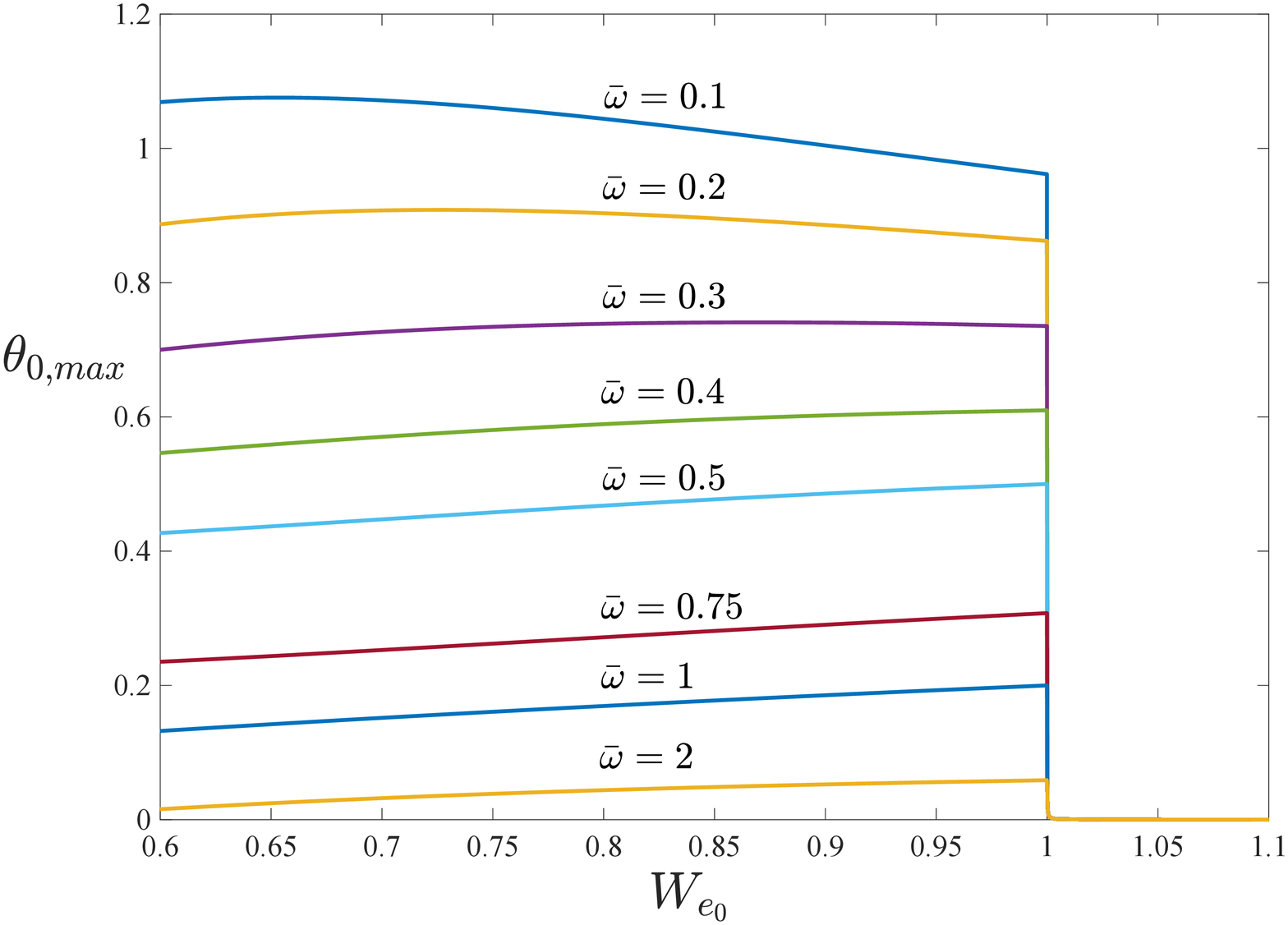}
  \caption{The maximum angle at the slot, $\theta_{0,\text{max}} =  \protect\underaccent{\forall t}{\text{max}}[\arctan(\text{Re}(\partial\bar{F}/\partial x))]_{\bar{x}=0}$, as a function of slot Weber number, $W_{e_0}$, for a range of dimensionless disturbance frequencies, $\bar{\omega}$, obtained from the solution of (\ref{ODE}) with $\beta=1$. The maximum angle at the slot, $\theta_{0,max}$, is taken here to correspond with the dimensionless curtains shown in figures 2-9 (as well as additional cases).}
  \label{end}
\end{figure} 

\clearpage
\appendix
\section{}
\subsection{Power Series} \label{AppendixPower}
 For subcritical cases ($W_{e_0} <1$), the power series $\Bar{H}(z) = \sum\limits_{j=0}^{\infty} (\lambda_j + i \Gamma_j) z^j$, centered at $z = 0$ (the critical point), is a solution of (\ref{ODE}) on the interval $-b < z < b$, if the following conditions are met, where $b = c + 1$ and $Q_{j} = [\lambda_j, \Gamma_j ]^{T}$:

\begin{equation} \label{Q1}
  Q_{1} = \left(A_1Q_0 + \left[ \begin{array} {cc}
b^4 \\ 
-2\bar{\omega}b^5 \\
\end{array} \right]\right)s_1
  \end{equation} 
\begin{equation} \label{Q2}
  Q_{2} = \left(A_2Q_1+B_2Q_0 + \left[ \begin{array} {cc}
12b^3 \\ 
-12b^4\bar{\omega} \\
\end{array} \right]\right)s_2
\end{equation} 
\begin{equation} \label{recursion}
  Q_{3} = \left(A_{3}Q_2+B_{3}Q_{1}+C_{3}Q_{0} + \left[ \begin{array} {cc}
27b^2 \\ 
-18b^3\bar{\omega} \\
\end{array} \right]\right)s_3
  \end{equation} 
  \begin{equation} \label{recursion}
  Q_{4} = \left(A_{4}Q_3+B_{4}Q_{2}+C_{4}Q_{1} + \left[ \begin{array} {cc}
16b \\ 
-8b^2\bar{\omega} \\
\end{array} \right]\right)s_4
  \end{equation}
 And for all $j>3$
\begin{equation} \label{recursion}
  Q_{j+1} = (A_{j+1}Q_j+B_{j+1}Q_{j-1}+C_{j+1}Q_{j-2})s_{j+1}
  \end{equation} 

 where:
\begin{equation}
\setlength{\arraycolsep}{3pt}
\renewcommand{\arraystretch}{2.5}
\label{M1}
A_j = \left[
\begin{array}{cc}
 A_{11j}& A_{12j} \\
 A_{21j}  & A_{22j}
\end{array} \right]
\end{equation} 

\begin{equation}
A_{11j} = b(j+1)^2(b^2\bar{\omega}^2-j(j-1))-8b^3\bar{\omega}^2j(j+1)
\end{equation}

\begin{equation}
A_{12j} = 2\bar{\omega} b^2(2j(j+1)^2+(j+1)(b^2\bar{\omega}^2-j(j-1))
\end{equation}

\begin{equation}
A_{21j} = -2\bar{\omega} b^2(2j(j+1)^2+(j+1)(b^2\bar{\omega}^2-j(j-1))
\end{equation}

\begin{equation}
A_{22j} = b(j+1)^2(b^2\bar{\omega}^2-j(j-1))-8b^3\bar{\omega}^2j(j+1)
\end{equation}

\begin{equation}
\setlength{\arraycolsep}{3pt}
\renewcommand{\arraystretch}{2.5}
\label{M2}
B_j = \left[
\begin{array}{cc}
 B_{11j} & B_{12j} \\
 B_{21j} & B_{22j} \\
\end{array} \right]
\end{equation} 

\begin{equation}
B_{11j} = 2b^2\bar{\omega}^2(j+1)(3-j)
\end{equation}

\begin{equation}
B_{12j} = 2b\bar{\omega}(j+1)(2b^2\bar{\omega}^2+(j+1)(j-1))
\end{equation}

\begin{equation}
B_{21j} = -2b\bar{\omega}(j+1)(2b^2\bar{\omega}^2+(j+1)(j-1))
\end{equation}

\begin{equation}
B_{22j} = 2b^2\bar{\omega}^2(j+1)(3-j)
\end{equation}

\begin{equation} 
\setlength{\arraycolsep}{3pt}
\renewcommand{\arraystretch}{2}
\label{M3}
C_j= \left[ \begin{array} {cc}
C_{11j} & C_{12j} \\ 
C_{21j}  & C_{22j} \\
\end{array} \right]
\end{equation} 

\begin{equation}
C_{11j} = b\bar{\omega}^2(j+1)^2
\end{equation}

\begin{equation}
C_{12j} = 2b^2\bar{\omega}^3(j+1)
\end{equation}

\begin{equation}
C_{21j} = -2b^2\bar{\omega}^3(j+1)
\end{equation}

\begin{equation}
C_{22j} = b\bar{\omega}^2(j+1)^2
\end{equation}

\begin{equation}
  s_j = \frac{1}{b^2j^4+4b^4\bar{\omega}^2j^2}\\
\end{equation}
  
 If $\lambda_0$ and $\Gamma_0$ are known, the rest of the parameters are determined. So we have a two-parameter family of solutions. In particular cases the two parameters are determined by the boundary condition, which is the Dirichlet condition given by Equation (2.13) for the cases we consider in this paper.

\subsection{Power Series Convergence Proof} \label{AppendixConvergence}
 Here we prove that the power series $\Bar{H}(z) = \sum\limits_{j=0}^{\infty} (\lambda_j + i \Gamma_j ) z^j$ converges for $|z| < b$ . Let $T_j = \Vert Q_j \Vert$. It follows from Equation A5 and the triangle inequality that 

\begin{equation}T_{j+1} \leq \Vert A_j s_j \Vert T_j + \Vert B_j s_j\Vert T_{j-1} + \Vert C_j s_j \Vert T_{j-2} \end{equation}
  
 The diagonal elements of $A_j$ are fourth-order polynomials in $j$ with leading coefficient $b$, the off-diagonal elements are third-order polynomials in $j$, and $s_j$ is a fourth-order polynomial with leading coefficient $b^2$. Thus the matrix $A_j s_j$ approaches $\frac{1}{b}I$, where $I$ is the identity matrix, as $j$ goes to infinity. All of the elements in the matrices $B_j$ and $C_j$ are polynomials of order less than four, so $B_j s_j$ and $C_j s_j$ approach the zero matrix as $j$ approaches infinity. Thus for every $r > 0$, there exists an integer $K$, such that for all $j > K$:

\begin{equation}
  \Vert A_j s_j \Vert < \frac{1}{b} +r \end{equation}
\begin{equation}\Vert B_j s_j \Vert < \frac{r}{b}\end{equation}
\begin{equation}\Vert C_j s_j \Vert < \frac{r}{b^2}\end{equation}

 If $\gamma > \left(\frac{1}{b} + 3 r \right)$, then: 
\begin{equation}
  \gamma > \frac{1}{b} + r + \frac{r}{\gamma b}+ \frac{r}{\gamma^2 b^2}
\end{equation}
 Let $q > 0$ be such that $T_j < q \gamma^j$, for all $j < K+3 $.
 
 Then,

\begin{equation}q \gamma^{j} > q\gamma^{j-1}(\frac{1}{b} + r) + q\gamma^{j-2}\frac{r}{ b}+ q\gamma^{j-3}\frac{r}{b^2} > (\frac{1}{b} + r) T_{j-1} + \frac{r}{b}T_{j-2} + \frac{r}{b^2} T_{j-3} \geq T_{j}\end{equation}

 Therefore by induction, $T_j < q \gamma ^j$ for all $j$. So $\sum\limits_{j=0}^{\infty} T_j|z^j|$ is dominated by the geometric series $\sum\limits_{j=0}^{\infty} q \gamma^j|z^j|$ and this converges for $|z| < \frac{1}{\gamma}$. Therefore, $\sum\limits_{j=0}^{\infty} Q_j z^j $ converges absolutely for $|z|<\frac{1}{\gamma}$. Because we have the derived bound for every $r$, where $\gamma = r+\frac{1}{b}$, we can take $r$ as small as we'd like to and establish that $b$ is the radius of convergence.

\newpage
\bibliographystyle{plain}

 \end{document}